\documentclass{sig-alternate-05-2015}
\usepackage{subcaption}
\usepackage{subfig}
\usepackage{multirow}
\usepackage{soul}
\usepackage{color}
\usepackage[colorinlistoftodos]{todonotes}
\begin{document}

\CopyrightYear{2016} 
\setcopyright{acmcopyright}
\conferenceinfo{CIKM'16 ,}{October   24-November  28, 2016, Indianapolis, IN, USA}
\isbn{978-1-4503-4073-1/16/10}\acmPrice{\$15.00}
\doi{http://dx.doi.org/10.1145/2983323.2983675}

\clubpenalty=10000 
\widowpenalty = 10000





%

\title{Anomalies in the peer-review system:\\ A case study of the journal of High Energy Physics}
%
%
%
%
%

\numberofauthors{4} 
%
 \author{
\alignauthor
Sandipan Sikdar\\
       \affaddr{Dept. of CSE}\\
       \affaddr{IIT Kharagpur}\\
       \affaddr{West Bengal, India -- 721302}\\
       \email{sandipansikdar@cse\\.iitkgp.ernet.in}
\alignauthor
Matteo Marsili\\
       \affaddr{ICTP}\\
       \affaddr{Strada Costiera}\\
       \affaddr{34014 Trieste, Italy}\\
       \email{marsili@ictp.it}
 \and
\alignauthor Niloy Ganguly\\
       \affaddr{Dept. of CSE}\\
       \affaddr{IIT Kharagpur}\\
       \affaddr{West Bengal, India -- 721302}\\
       \email{niloy@cse.iitkgp.ernet.in}
\alignauthor Animesh Mukherjee\\
       \affaddr{Dept. of CSE}\\
       \affaddr{IIT Kharagpur}\\
       \affaddr{West Bengal, India -- 721302}\\
       \email{animeshm@cse.iitkgp.ernet.in}
 }

\maketitle
\begin{abstract}
Peer-review system has long been relied upon for bringing quality research to the notice of the scientific community and also preventing flawed research from entering into the literature. The need for the peer-review system has often been debated as in numerous cases it has failed in its task and in most of these cases editors and the reviewers were thought to be responsible for not being able to correctly judge the quality of the work. This raises a question ``Can the peer-review system be improved?'' Since editors and reviewers are the most important pillars of a reviewing system, we in this work, attempt to address a related question - given the editing/reviewing history of the editors or reviewers ``can we identify the under-performing ones?'', with citations received by the edited/reviewed papers being used as proxy for quantifying performance. We term such reviewers and editors as anomalous and we believe identifying and removing them shall improve the performance of the peer-review system.   
Using a massive dataset of Journal of High Energy Physics (JHEP) consisting of $29k$ papers submitted between $1997$ and $2015$ with $95$ editors and $4035$ reviewers and their review history, we identify several factors which point to anomalous behavior of referees and editors.
In fact the anomalous editors and reviewers account for $26.8\%$ and $14.5\%$ of the total editors and reviewers respectively and for most of these anomalous reviewers the performance degrades alarmingly over time. 

\end{abstract}


\section{Introduction}
\label{introduction}

Before the contributions of a paper are brought to the notice of the research community, it has to usually pass through a peer-review process, whereby, the correctness and the novelty of the paper is judged by a set of knowledgeable peers. The primary intent of which is to prevent flawed research from getting into mainstream literature \cite{kassirer1994peer}. 

\noindent{\bf Debates on peer-review system:}
The effectiveness of this system has been put to question in numerous cases (\cite{ingelfinger1974peer,relman1989good,smith2006peer}) with flawed research being added to literature while significantly novel contributions being rejected. That the reviewers often fail to reach consensus (\cite{cole1981chance}) and that rejected papers are often cited more in the long run (\cite{braatz2014papers}), have already been pointed out. Although there have been several proposals to make it more effective (\cite{caswellimproving,graffy2006improving,mcnutt1990effects}), 
the research community is coming to a conclusion that although peer-review system is indispensable it is nonetheless flawed \cite{bacchetti2002peer}. 

\noindent {\bf Entities in the peer-review system:} The effectiveness of the peer-review system is dependent directly on the knowledge and training of the editors and reviewers. The editor is responsible for identifying the correct set of referees who can give expert comments on the submission and also for taking the final decision whether a particular paper should be accepted or rejected. The assisting reviewers send their views on the paper in the form of a report. This report is an important part of the whole process as it not only forms the basis of the acceptance/rejection decision but is also sent to the authors for further improvement of the paper.  

\noindent{\bf Anomalous behavior:} 
Ideally impactful papers should be accepted for publication while flawed works should be rejected. We quantify the impact of a paper by the citations it garnered. Thus, a paper getting accepted but managing to garner very less or no citation should be attributed to the anomaly of the system; similarly, a paper getting rejected by the peer-review-system but garnering large number of citations in the long run is also an anomaly. 
We in this paper investigate the reasons behind the anomalous behaviors (\cite{chandola2009anomaly}) of the reviewers and editors as they are the most important entities of the peer-review system. 
Note that although the number of such anomalous editors or referees might be small compared to the number of normal editors or reviewers  (as is usually the case with any anomalous set), the damage they can cause to the peer-review system could be irreparable and therefore a thorough investigation of this set is extremely necessary. 

\noindent{\bf Characterizing anomalous editors and reviewers:} A thorough investigation of the behavior of the {\bf editors} shows that those editors who (i) are assigned papers more frequently, (ii) select reviewers from a very small set, (iii) assign themselves as reviewers more often (rather than assigning other reviewers) are often under-performers and hence anomalous.  
Similarly, for {\bf reviewers} we observe the following behaviors to be anomalous - (i) frequent assignments, (ii) very small or very large delay in sending reports, (iii) reviewing papers in very specific topics, (iv) assignments from a very small set of editors or in some cases a single editor, (v) very high or very low proportion of acceptance,  (vi) large delay in informing the editor about inability to review and (vii) often declining to review. Papers accepted by reviewers with such behaviors are often low cited while those rejected by them are often highly cited.

\noindent{\bf Identifying anomalous editors and reviewers:} All the above observations lead us to believe that anomalous editors and reviewers can be differentiated from the genuine contributors. To this aim we use these observations as features and by leveraging anomaly detection techniques we are indeed able to filter out the anomalous editors and reviewers. In specific we use $k$-means clustering \cite{hartigan1979algorithm} to classify normal and anomalous editors and reviewers.
We find $26.8\%$ of the editors and $14.5\%$ of the reviewers to be anomalous.
We further observe that the papers accepted by these anomalous reviewers are on average cited less while those rejected by them are cited more. 


\noindent{\bf Organization of the paper:} The rest of the paper is organized as follows. In section~\ref{dataset} we describe in detail the dataset we used for our analysis and point out certain important features. In section~\ref{anomalies} we identify several factors which help in characterizing anomalous editors and referees. In section~\ref{prediction} we identify anomalous editors and reviewers. 
We further assess the performance of the anomalous reviewers in section~\ref{profile}. 
We finally conclude in section~\ref{conclusion} by highlighting our main contributions and pointing to certain future directions.

\section{Dataset}
\label{dataset}

As mentioned earlier, the main aim of this work is to understand the anomalous behaviors of the peer-review system. For this purpose, we use the dataset provided to us by the Journal of High Energy Physics (JHEP)\footnote{jhep.sissa.it/jhep}. JHEP is one of the leading journals with an impact factor of $6.1$ \footnote{http://www.springer.com/physics/particleandnuclear \\ physics/journal/13130} (2014) and publishes theoretical, experimental and phenomenological papers. 

The dataset consists of {\bf 28871} papers that were submitted between 1997 (year of inception) and 2015 of which {\bf 20384} were accepted and {\bf 7073} were rejected. The rest of the papers were either withdrawn by the authors or the final decisions were not available. 
The meta information available for each paper are (i) title, (ii) name of the contributing authors, (iii) abstract, (iv) date of publication and (v) number of citations till 2015. More importantly, we have for each paper full action history from the date of submission to the date of publication including the editor(s) and the reviewer(s) involved for the review of the paper, the report provided by the reviewer and the report sent to the authors by the editor. 
We further queried the {\em Inspire} \footnote{https://inspirehep.net/} database to obtain the meta information of the papers (not published/rejected in JHEP). Using this information we created a citation profile for each paper i.e., citations received by the paper per year from the year of its publication. Garfield et. al. ~\cite{garfield1999journal} had noted that most papers receive the bulk of their citations within the first three years of publication. Moreover it is known that old papers generally have more citations, as the paper had more time to accumulate the citations. Thus to account for this effect,  we  calculated total citations received by each paper in the first three years from its year of publication (e.g., for a paper published in 2007 we consider the citations received by it till 2010). For the rest of this article, by citation of a paper we refer to the number of citations it received in the first three years from the year of its publication and we only consider the papers published between 1997 to 2012 for our experiments. 
Some general properties of the whole dataset are summarized below - \\
(i) Number of unique editors in the dataset is 95 while the number of reviewers is 4035. 
(ii) There are $15127$ unique authors in the dataset and of those $12434$ have at least one accepted paper.
(iii) Average number of submissions per author is $5.18$ while the average number of authors per paper is 2.87.
(iv) Average number of reviews for accepted and rejected are $1.76$ and $1.35$ respectively.
(v) Average number of assignments per editor is $298.28$ while per reviewer it is $7.52$.

\section{Anomalous behavior}
\label{anomalies}
In the peer-review process each submission is assigned to an editor who in turn assigns one or more reviewers with the task of judging the quality of the contributions of the submitted paper. The reviewer submits a report to the editor who in turn takes the final decision as to accept or reject the paper based on the report. Therefore, the editors and the reviewers are the two important entities of the peer-review system and they are mainly responsible for ensuring that flawed research does not get into the literature while at the same time correctly identify impactful contributions for publication.  
 So in our setting we define the following two cases to be anomalous - \\
(i) Accepted papers having low citation (research wrongly judged as impactful). \\
(ii) Rejected papers having high citation (quality research wrongly judged as flawed). \\
In this section we look into the anomalous behavior of the two important entities of the peer-review process: (i) the editors and (ii) the reviewers.

\subsection{Editor}
\label{editor}   

\begin{figure}
\centering
\includegraphics[width=.48\textwidth]{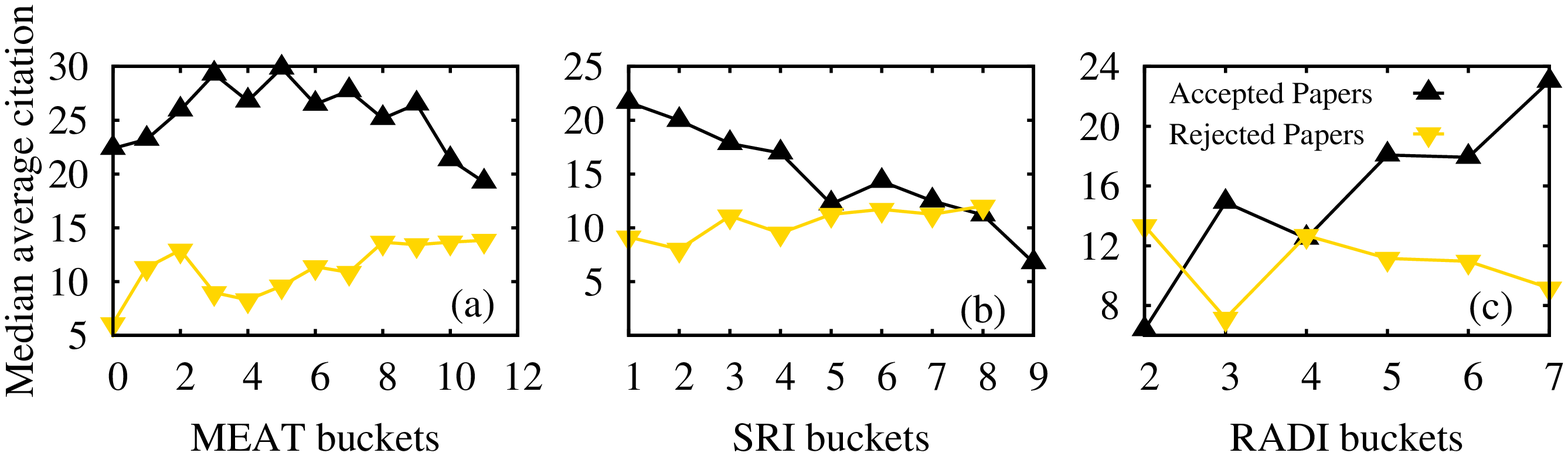}
\caption{\label{fig3}(a) Median Average citation (MAC) versus $MEAT$. $MEAT$ values are bucketed into 12 bins of equal size with range(1, 498.8).(b) MAC versus $SRI$ and (c) MAC versus $RADI$. For both (b) and (c), the x-axis values are bucketed by values corresponding to ($\geq$ 0 and $<$ 0.1), ($\geq$ 0.1 and $<$ 0.2) and so on.}
\end{figure}

\begin{figure}[!ht]
\centering
\includegraphics[scale=0.27]{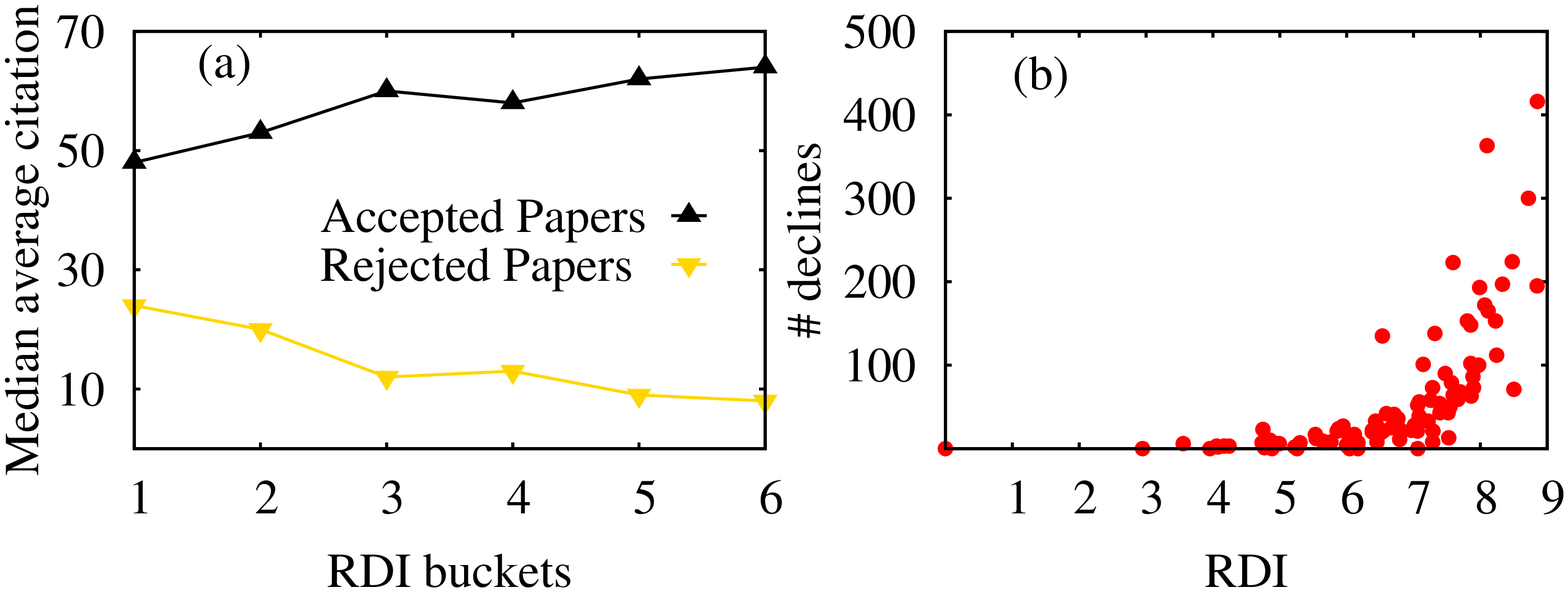}
\caption{\label{fig_sri} (a) Median Average citation versus $SRI$. $SRI$ values are bucketed by values corresponding to ($\geq$ 0 and $<$ 0.1), ($\geq$ 0.1 and $<$ 0.2) and so on. (b) $RDI$ versus number of declines. Increasing trend indicates higher the $RDI$, higher is the number of declines.}
\end{figure}

\begin{figure}[!ht]
\centering
\includegraphics[width=0.48\textwidth]{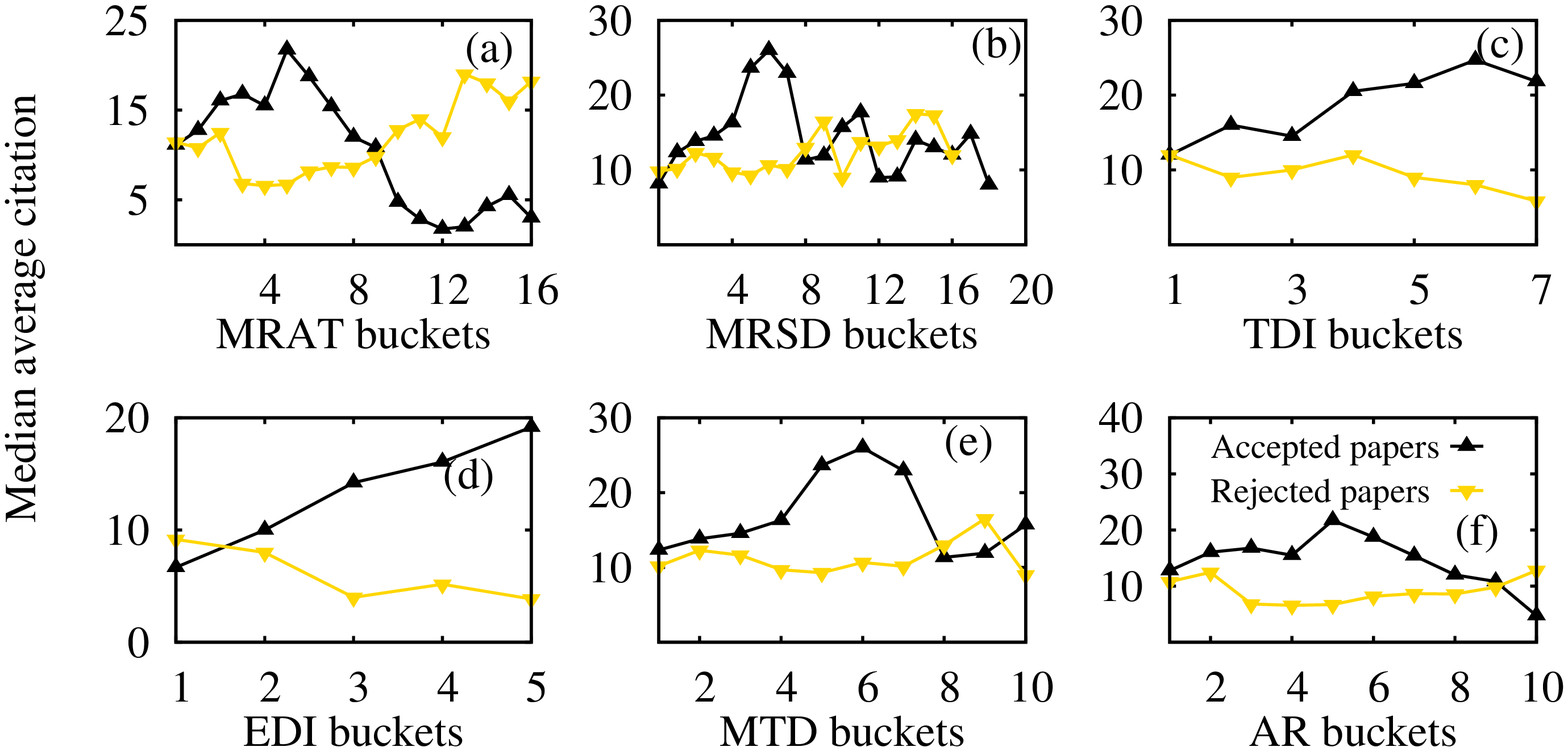}
\caption{\label{fig5}(a) Median Average citation (MAC) versus $MRAT$. $MRAT$ values are bucketed into 20 buckets of equal size with range(1,498.8),(b) MAC versus $MRSD$ (c) MAC versus $TDI$, (d) MAC versus $EDI$, (e) MAC versus $MTD$ and (f) MAC versus $AR$. For both (c),(d) and (e), the x-axis values are bucketed by values corresponding to ($\geq$ 0 and $<$ 0.1), ($\geq$ 0.1 and $<$ 0.2) and so on. For (b) and (f) values (x-axis) are divided into 10 buckets of equal size.}
\end{figure}

We begin by analyzing the anomalous behavior of the editors. We define the behavior of an editor to be anomalous if the papers assigned to her are on average cited less when accepted or are cited more when rejected. In specific, we investigate different factors related to the editor that can lead to such anomaly.
\subsubsection{Mean Editor Assignment Time (MEAT)}
For each editor we obtain the time span (in days) between any two consecutive assignments and calculate the average time span between the two assignments. 
Formally, we define for editor $i$, $MEAT_{i}$ as
\begin{center}
$MEAT_{i}=\frac{1}{n-1}\sum (\delta_{j+1} - \delta_{j})$
\end{center}
where $n$ is the total number of assignments to the editor $i$ and $\delta_{j}$ is the date of the $j$$^\textrm{th}$ assignment. 
In figure~\ref{fig3}(a) we bin the editors based on the $MEAT$ and calculate the median average citation of the papers assigned to the editors in each bin. 
{We observe that for accepted papers very low or very high $MEAT$ values lead to lower citations. An exact opposite behavior is observed for rejected papers. 
This indicates that editors who are assigned time and again (low $MEAT$) or rarely (high $MEAT$) often fail to judge the quality of the papers assigned to them.} 

\subsubsection{Self Review Index (SRI)}

{\bf Self Review Index (SRI)} measures the fraction of papers for which the editor assigned herself as the reviewer. 
Formally, for an editor $i$, we define $SRI_{i}$ as 
\begin{center}
$SRI_{i}=\frac{\varrho_{i}}{\rho_{i}}$
\end{center}
where $\rho_{i}$ is the number of papers $i$ was assigned as editor while $\varrho_{i}$ is the number of papers $i$ assigned herself as reviewer. We observe that with increasing values of $SRI$ the median average citation for accepted papers decreases while that for rejected papers increases (refer to figure \ref{fig3}(b)). 

\subsubsection{Referee-Author pair Diversity Index (RADI)}

We observe that editors in numerous cases assign papers from a certain author to only a certain reviewer. To investigate whether this allows for less impactful research from this author getting accepted, we define a metric which we call {\bf Referee-Author pair Diversity Index (RADI)}. Formally we define for editor $i$, the $RADI_{i}$ score as 

\begin{center}
$RADI_{i}=-\sum \limits_{j,k} p_{j,k} \log p_{j,k}$
\end{center}

where $p_{j,k}$ denotes the proportion of times a paper from author $k$ was assigned to reviewer $j$ by the editor $i$. In figure~\ref{fig3}(c) we bin the editors based on the $RADI$ and calculate the median average citation of the papers assigned to the editors in each bin. We observe that more the diversity score higher is the citation of the accepted papers and correspondingly lower is the citation of the rejected papers.


\subsubsection{Referee Diversity Index (RDI)}
As a following step, we check whether an editor always chooses from a fixed set of reviewers or a diverse set of reviewers while making a paper assignment and, more importantly, does this influence the performance of the editor in terms of the impact of the reviewed paper. We define for each editor($i$) a metric called {\bf Referee Diversity Index ($RDI_{i}$)} as -  
\begin{center}
$RDI_{i}=-\sum \limits_{j} p_{j}\log p_{j}$
\end{center}
where $p_{j}$ denotes the proportion of times reviewer $j$ was assigned a paper by editor $i$. More diverse the set of reviewers higher is the score. In figure~\ref{fig3}(b) we bin the editors based on the $RDI$ and calculate the median average citation of the papers assigned to the editors in each bin. We observe that more the diversity score, higher is the citation of the accepted papers and correspondingly lower is the citation of the rejected papers.

A summary statistic of all the above factors that may be used to identify anomalous editors are noted in Table~\ref{summary_stat}.

The dataset allows us to find out the cases when the reviewer declined to review a paper on being assigned by an editor. We observe that editors with high $RDI$ are also declined more often. In figure~\ref{fig_sri}(b) we plot $RDI$ value and the number of declines for each editor. An increasing trend indicates that more diversely the editor tries to select reviewers more she gets declined by the reviewers. This in many cases may force the editor to be less proactive and always select from a specific set of `reliable' referees.

\subsection{Reviewer}
\label{reviewer}

In this section, we investigate anomalous behavior of the referees. Recall that we define the behavior of a reviewer to be anomalous if the papers accepted by her are low cited or the papers rejected by her are highly cited. As in case of the editors, here also we investigate different factors that could be indicative of such anomalous behavior.  

\subsubsection{Mean Reviewer Assignment Time (MRAT)}

This is essentially same as MEAT. For a reviewer $i$, we define $MRAT_{i}$ as
\begin{center}
$MRAT_{i}=\frac{1}{n-1}\sum (\delta_{j+1} - \delta_{j})$
\end{center}
\noindent where $n$ is the total number of assignments of reviewer $i$ and $\delta_{j}$ is the date of the $j^\textrm{th}$ assignment. In figure~\ref{fig5}(a) we plot $MRAT$ (binned) and median average citation of the papers reviewed for each reviewer. We observe that papers reviewed by reviewers with low $MRAT$ (high frequency of assignment) tend to be cited less and increases as $MRAT$ increases. This is followed by again a steep decrease in citation. This indicates that the reviewers assigned very frequently are often less reliable while those assigned only occasionally are also not likely to correctly judge the quality of the paper.  

\subsubsection{Mean Report Sending Delay (MRSD)}

We argue that the time taken by a reviewer to send back the review report could be an indicator of his performance. If a reviewer on average sends back the review very quickly it is highly likely that the review was done in a hurry. Similarly, if the report was sent after being reminded by the editor numerous times, it is also highly likely the review report could be anomalous. For a reviewer we calculate the time delay between the date of her assignment and the date she sent back the report for each of her assignments. To measure $MRSD$ we calculate the mean value of all the delays. Note that we do not consider the assignments which the reviewer declined. Formally, for a reviewer $i$, we define $MRSD_{i}$ as 

\begin{center}
$MRSD_{i}=\frac{1}{n}\sum(\delta_{i}-\Delta_{i})$
\end{center}

where $n$ is the total number of assignments, $\Delta_{i}$ is the date of assignment and $\delta_{i}$ is the date when the report was received by the editor. On plotting against median average citation we observe a similar trend as was observed in case of $MRAT$ (refer to figure~\ref{fig5}(b)). Papers reviewed by reviewers with low $MRSD$ value are often less cited, indicating that reviewers sending back their report very quickly often do it in a hurry and fail to correctly judge the quality of the paper while those taking very long to send report are prone to failure as well. 

\subsubsection{Topic Diversity Index (TDI)}

JHEP associates with each submission a set of keywords which roughly indicates the domain of the work. We use these associated keywords as a proxy for topic. For each reviewer, we segregate all the keywords of the papers reviewed by her which we call the keyword corpus for the reviewer. Formally for a reviewer $i$, we define $TDI_{i}$ as 

\begin{center}
$TDI_{i}=-\sum \limits_{j} p_{j}\log p_{j}$
\end{center}

\noindent where $p_{j}$ is the proportion of keyword $j$ in the keyword corpus for reviewer $i$. We segregate the reviewers based on the diversity score and calculate the median average citation of the papers reviewed by them. We observe that the median average citation 
for reviewers with low $TDI$ are low mainly because the number of papers reviewed by them are also less. The value increases with  increasing $TDI$ (refer to figure~\ref{fig5}(c)). The reviewers with low $TDI$ are often the ones who have reviewed a very small number of papers while the reviewers with high $TDI$ are mostly assigned papers by a large number of editors.

\subsubsection{Editor Diversity Index (EDI)}

Reviewers could be selected for review by a large set of editors or could only be selected by a single or a small set of editors. We check whether a reviewer selected by many editors is more reliable compared to one who is selected by a single or a very small set of editors. To this aim we assign each reviewer a score called Editor Diversity Index, $EDI_{i}$ which is defined as 

\begin{center}
$EDI_{i}=-\sum \limits_{j} p_{j}\log p_{j}$
\end{center}

where $p_{j}$ represents the proportion of times reviewer $i$ was assigned by editor $j$. We segregate the reviewers based on $EDI$ and calculate the median average citation of the papers reviewed by them. We observe that as $EDI$ increases median average citation also increases (refer to figure \ref{fig5}(d)) indicating that reviewers assigned by multiple editors are often more reliable.

\begin{table}[htpb]
\centering
\caption{Features used for detecting anomalies.}
\label{summary_stat}
\small
\begin{tabular}{|l|l|l|l|l|l|l|}
\hline
                        & Factor                                     & Mean & Median & Max & Min & \begin{tabular}[c]{@{}l@{}}St.\\ Dev\end{tabular} \\ \hline
\multirow{3}{*}{Editor} & $MEAT$         & 35.06     &  29.1      &  108.25   &  3.28   & 23.19                                                             \\  
                        & $RDI$              & 6.57     &  6.79      & 8.85    & 0.0    &  1.44                                                            \\ 
                        & $RADI$ & 8.86     & 9.21       & 11.94    & 0.0    &  1.87 
                         \\ 
                        & $SRI$ & 0.28     & 0.25       & 0.85    & 0.0    &  0.19                                                           
                        \\ \hline
\multirow{6}{*}{Reviewer} & $MRAT$       & 363.3     & 193.7       & 5389    & 26.9    & 508.9                                                             \\  
                        & $MRSD$           & 19.28     & 17.50       & 122    & 16.5    &  11.45                                                            \\  
                        & $TDI$                &  4.07    &  3.96      & 8.10    & 1.0    &  1.44                                                            \\  
                        & $EDI$               &  1.12    &   0.91     &  4.58   & 0.0    &   1.19                                                           \\  
                        & $AR$                      &  0.65    &   0.71     &   1.0  & 0.0    &    0.2                                                          \\  
                        & $MTD$                 &   3.86   &  3.12      & 69.0    & 1.0    & 4.96                                                             \\ 
                        & $DFI$                 & 0.19     & 0.12       & 1.0    & 0.0    & 0.30                                                             \\ \hline
                        
\end{tabular}
\end{table}

\subsubsection{Mean Time to Decline (MTD)}

We further investigated the cases where the reviewer declined the assignment. In specific, we calculated the time delay (in days) 
between the date she was assigned and the date she conveyed her decision of declining to review. For each reviewer we define {\bf Mean Time to Delay}, $MTD_{i}$ as 

\begin{center}
$MTD_{i}= \frac{1}{d}\sum \limits_{j}(\mu_{j} - \Delta_{j})$
\end{center}

where $d$ is the number of assignments that reviewer $i$ declined and $\mu_{j}$ and $\Delta_{j}$ are respectively the date of assignments and date of reply for paper $j$ by reviewer $i$. We segregate the reviewers based on their $MTD$ values and calculate the median average citation. We observe that the reviewers who delay often in reporting their decision to the editor of being unable to review usually tend to fail in judging a paper quality when they do review (refer to figure~\ref{fig5}(f)).

\subsubsection{Acceptance Ratio (AR)}
Acceptance Ratio ($AR$) of a reviewer is defined as the proportion of papers accepted by the reviewer. For a reviewer $i$, $AR_{i}$ is formally defined as 

\begin{center}
$AR_{i}=\frac{a_{i}}{a_{i}+r_{i}}$
\end{center}

\noindent where $a_{i}$ and $r_{i}$ respectively denote the number of papers accepted and rejected by reviewer $i$. We observe that reviewers with high $AR$ often accept  less impactful papers while reviewers with very low $AR$ often fail to identify quality research (refer to figure~\ref{fig5}(e)). Note that the reviewers are segregated based on their respective $AR$ values while the median average citation is calculated. They are segregated into bins based on the $AR$ values where typically the bins are ($\geq$ 0 and $< 0.1$), ($\geq$ 0.1 and $<$ 0.2) and so on.

\begin{figure}
\centering
\includegraphics[scale=0.3]{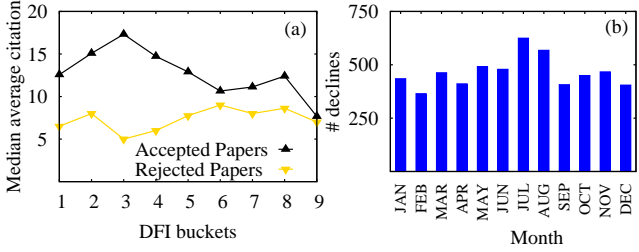}
\caption{\label{fig_dfi}(a) Median Average citation versus $DFI$. $DFI$ values are bucketed by values corresponding to ($\geq$ 0 and $<$ 0.1), ($\geq$ 0.1 and $<$ 0.2) and so on. (b) Number of declines versus the month of the year.}
\vspace{-.4cm}
\end{figure}

\subsubsection{Decline Fraction Index (DFI)}

{\bf Decline Fraction Index (DFI)} for a reviewer is the fraction of times she declined to review. For a reviewer $i$, we define $DFI_{i}$ as
\begin{center}
$DFI_{i}=\frac{\vartheta_{i}}{\theta_{i}}$
\end{center}

\noindent where $\theta_{i}$ is the total number of assignments while $\vartheta_{i}$ is the number of times $i$ declined to review. 
In figure \ref{fig_dfi}(a) we plot median average citation versus $DFI$. We observe that for accepted papers the citation is higher for lower $DFI$ values and it drops as $DFI$ increases indicating that reviewers declining too frequently often fail to judge the quality of the paper assigned to them.

A summary statistics of all the above factors that may be used to identify anomalous referees are noted in Table~\ref{summary_stat}.

We further looked into the data and made some interesting observations which are summarized below - \\
(i) A bulk of the instances where a reviewer declined to review occurred in the month of July and August. This is represented in figure~\ref{fig_dfi}(b). This probably relates to the vacation time in the Europe and the US.
(ii) Of the 4035 reviewers 756 of the reviewers have not been assigned a paper for the last two years. On further investigation we observed that among these there are 505 such reviewers who in their immediate last review assignment agreed to review but did not send back the report.

\section{Identifying anomalous Editors and Reviewers}
\label{prediction}


In the previous sections we discussed how different factors indicate anomalous behavior of referees and editors. In this section, we check whether we can use them to automatically differentiate between normal and anomalous editors and referees. We propose separate unsupervised models for editors and reviewers.

\begin{figure}
\centering
\includegraphics[scale=0.27]{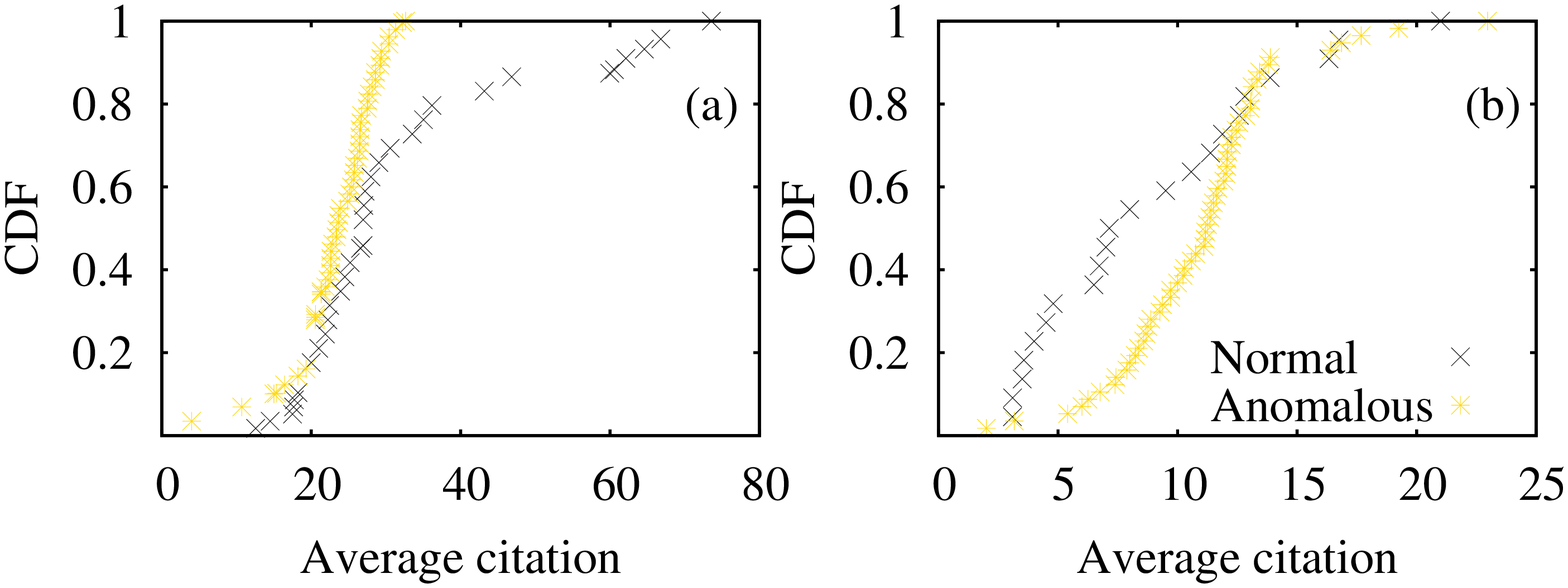}
\caption{\label{ed_pred}Cumulative distribution function of the average citations for the two sets of editors (anomalous and normal).} 
\vspace{-.4cm}
\end{figure}

\subsection{Editors}
For each editor $i$, we measure $MEAT_{i}$, $RDI_{i}$, $RADI_{i}$ and $SRI_{i}$ which form a feature vector. 
We also consider the editors who were assigned at least 5 papers and accepted at least 1 paper before 2013. 
To detect anomalies we use the 
$k-means$ clustering setting with $k=2$. The two clusters are of sizes 25 and 68 respectively. Clearly this set of 25 editors are the anomalous ones.
In figure \ref{ed_pred} we plot the cumulative distribution of average citation of accepted (figure \ref{ed_pred}(a)) and rejected  (figure \ref{ed_pred}(b)) papers. We observe that citation of accepted papers assigned to anomalous editors are significantly lower while citation of rejected papers are significantly higher compared to those assigned to normal editors.

\subsection{Reviewers}

\begin{figure}
\centering
\includegraphics[scale=0.27]{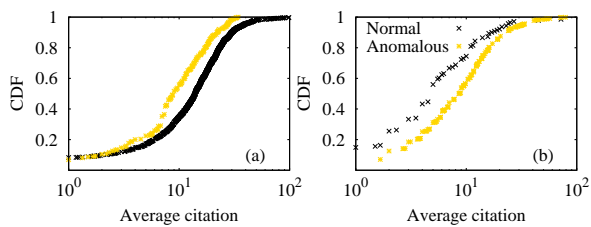}
\caption{\label{rev_pred}Cumulative distribution function of the average citations for the two sets of reviewers (anomalous and normal).}
\vspace{-.4cm}
\end{figure}

Similarly for each reviewer $i$ we associate a feature vector of size seven consisting of $MRAT_{i}$, $MRSD_{i}$, $TDI_{i}$, $EDI_{i}$, $AR_{i}$, $MTD_{i}$ and $DFI_{i}$. 
We filter out reviewers who have reviewed at least 5 papers and accepted at least 1 before 2013. This reduces our set of reviewers to 2328. By using $k-means$ clustering ($k=2$), we obtain two clusters of size 339 and 
1999. On plotting cumulative distribution of the average citation for accepted (refer to figure \ref{rev_pred}(a)) and rejected papers (refer to figure \ref{rev_pred}(b)), we observe that the papers accepted by anomalous reviewers are cited significantly lesser while those rejected by them are cited significantly higher compared to the normal referees.

\section{Profiling Anomalous Reviewers}
\label{profile}

\begin{figure}
\centering
\includegraphics[scale=0.3]{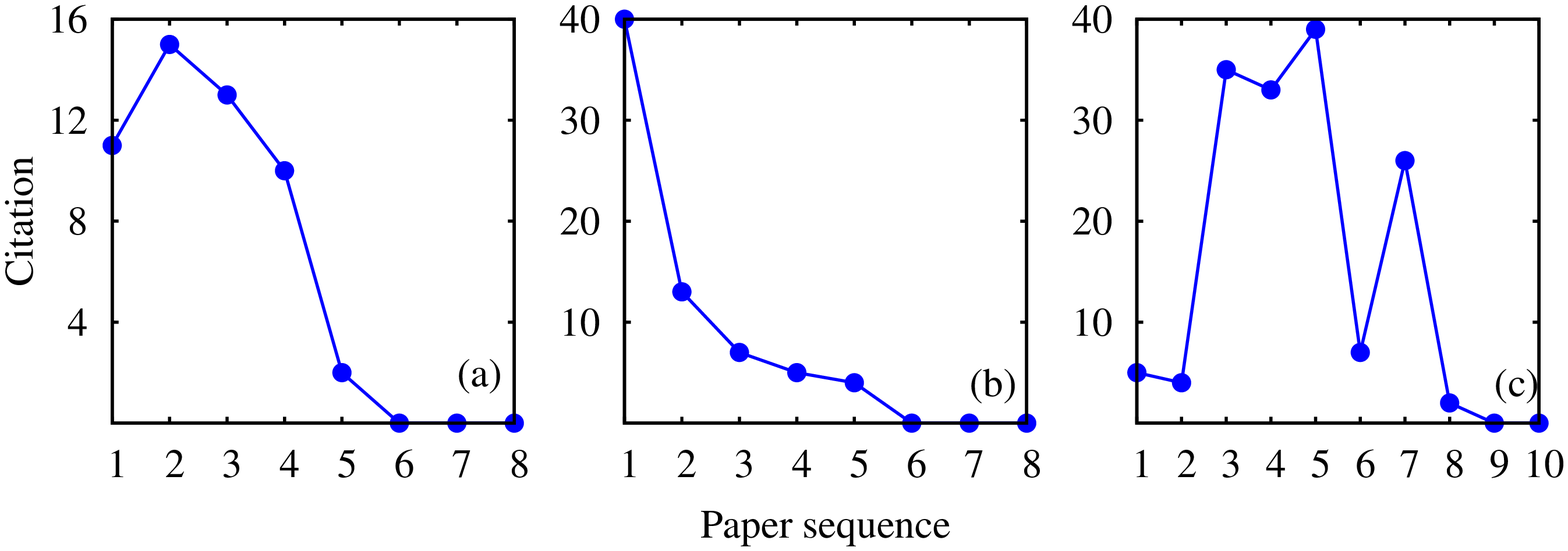}
\caption{\label{cit_prof} Mean citation profile of the reviewers in the three categories.}
\vspace{-.4cm}
\end{figure}

In this section we analyze in more details the performance of the anomalous reviewers. To this aim, we consider for each reviewer, the sequence of papers accepted by her and the citation accrued (within the first three years from publication) by each of these papers. A decreasing trend would suggest decline in performance of the reviewer. Depending on the trend we observe three broad categories within the set of anomalous reviewers \\
(i) performance deteriorates constantly over time (proportion = $42.5\%$, figure \ref{cit_prof}(a)).\\
(ii) performance is good for initial few papers but deteriorates in the long run (proportion = $22.6\%$, figure \ref{cit_prof}(b)).\\
(iii) performance fluctuates but has a deteriorating trend in the long run (proportion = $34.9\%$, figure \ref{cit_prof}(c)).

\section{Conclusion}
\label{conclusion}

In this paper we provided a framework for identifying anomalous reviewers and editors based on  their review history considering Journal of High Energy Physics as a case study. We identified several factors that are indicative of anomalous behavior of the editors as well as reviewers. In specific for editors we observed that - 
(i) high frequency of assignment, (ii) selecting from a very small set of referees for reviewing, (iii) assigning same reviewer to papers of same author and (iv) assigning herself as reviewer instead of assigning someone else could be indicative of anomalous behavior of the editor. 

Similarly for reviewers we observe that - 
(i) high frequency of assignment, (ii) delay in sending report, (iii) assignment from only a single editor or a very small set of editors (iv) 
very high or very low acceptance ratio and (vi) delay in notifying the editor about her decision to decline are also indicative of anomalous behavior and often leads to under-performance. Based on these factors we were able to identify anomalous referees and editors using an unsupervised clustering approach. 

\noindent{\bf Future directions:} We believe our findings could be useful in better assignment of editors and reviewers and thereby improve the performance of the peer-review system. Assigning good reviewers is an important part of the peer-review process and our findings allow for identifying under-performing referees. This could be a first step towards developing a reviewer-recommendation system whereby the editors are recommended a set of reviewers based on their performance. 
We plan to come up with such a system in subsequent works.

\section*{Acknowledgment}
We would like to thank publication team of Journal of High Energy Physics (JHEP) for providing us the necessary data and they were the only ones willing to provide it.


\end{document}